\documentclass[letterpaper]{article} 
\usepackage[preprint]{AAAI2027}  
\usepackage[hyphens]{url}  
\usepackage{graphicx} 
\usepackage{natbib}  
\usepackage{caption} 
\usepackage{algorithm}
\usepackage{algorithmic}

\usepackage{newfloat}
\usepackage{listings}

\usepackage{subcaption}
\usepackage{times}
\usepackage{helvet}
\usepackage{courier}
\usepackage[hyphens]{url}
\usepackage{natbib}
\usepackage{graphicx}
\usepackage{multirow}
\usepackage{booktabs}
\usepackage{microtype}
\usepackage{xcolor}
\usepackage{amsmath}
\usepackage{amssymb}

\DeclareCaptionStyle{ruled}{labelfont=normalfont,labelsep=colon,strut=off} 
\floatstyle{ruled}
\newfloat{listing}{tb}{lst}{}
\floatname{listing}{Listing}

\usepackage{booktabs}

\title{Capability-Routed Guard: Defending Large Reasoning Models Against Reasoning-Centric Jailbreaks}
\author{
    Yiyong Liu\textsuperscript{\rm 1},
    Yixin Wu\textsuperscript{\rm 1},
    Jun Sakuma\textsuperscript{\rm 2}
}
\affiliations{
    \textsuperscript{\rm 1}CISPA Helmholtz Center for Information Security\\
    \textsuperscript{\rm 2}Institute of Science Tokyo
}

\begin{document}

\maketitle

\begin{abstract}
Large reasoning models (LRMs) expose a new safety failure mode: adversarial prompts can manipulate reasoning context, task decomposition, or capability interpretation so that harmful objectives are processed as legitimate reasoning steps. Existing safeguards, including safety reminders, external classifiers, and self-checking wrappers, are often brittle because they either inspect the adversarial prompt directly or ask the target model to perform additional safety reasoning on the same surface that attacks exploit. We introduce \emph{Capability-Routed Guard} (CRG), a model-agnostic inference-time guardrail for closed-source LRMs, where defenders cannot inspect hidden reasoning traces or modify model weights. CRG reframes prompt defense as a capability-routing problem: a side-channel controller first constructs a trusted representation of the user's authorized task, active context, safety evidence, and capability-transfer risk, separating executable intent from untrusted reasoning context. This representation supports route-specific execution, allowing CRG to block high-risk requests, constrain ambiguous ones, and forward low-risk requests through trusted active context. Finally, CRG applies TraceCheck to verify consistency with the authorized task and invokes a restricted fallback to preserve utility for low-risk benign prompts. Extensive experiments demonstrate that CRG effectively mitigates diverse reasoning-centric jailbreaks while preserving benign utility and avoiding common over-refusal issues. Further analysis shows that its components contribute complementary benefits, highlighting the importance of coordinated defense mechanisms for securing large reasoning models.
\end{abstract}

\section{Introduction}
\label{sec:intro}
Large reasoning models (LRMs)~\cite{WWSBIXCLZ22,O24,D25} are increasingly used for tasks that require long-horizon planning, code synthesis, mathematical reasoning, scientific analysis, and autonomous decision support. Their strength comes from an expanded ability to decompose problems, maintain intermediate context, and reason over multi-step trajectories. This same capability, however, creates a new safety challenge. Unlike conventional chat models, where safety failures often arise from direct refusal bypasses, LRMs can be attacked through the reasoning process itself: adversarial prompts can manipulate task decomposition, inject false authority, dilute capability-transfer signals, or cause harmful objectives to appear as benign intermediate subtasks.

Recent jailbreaks~\cite{ZFSSB25,YTWWLLTW25,LJWZMW25,LFHXLX25,NZDHXWL26} show that this vulnerability is not limited to a single prompt pattern. Rather than merely asking the model to ignore its safety policy, these attacks increasingly shape the intermediate reasoning path: they pad harmful goals with benign-looking reasoning, guide the model through locally safe transformations, search automatically for effective reasoning wrappers, or embed unsafe objectives in elaborate fictional or encoded contexts. These strategies exploit the model's tendency to maintain coherence with the reasoning trajectory it has accepted, even when the final answer would transfer harmful capability. CoT-Hijacking~\cite{ZFSSB25} and SEAL~\cite{NZDHXWL26} are examples of this broader shift, but the underlying issue is more general than any individual attack template. We refer to attacks that exploit the reasoning trajectory, authority boundary, or capability interpretation of LRMs as \emph{reasoning-centric jailbreaks}.

Reasoning-centric jailbreaks expose limitations in many existing defense paradigms~\cite{IUCRIMTHFTK23,CXHL25,ZYKMWH24,LAJAYQ25,SHGLCYZH25,WZWHWYY26}. Most safeguards are designed to detect or suppress harmful prompts and responses, rather than to govern how a reasoning model interprets, routes, and executes a request. As a result, they often lack an explicit mechanism for separating trusted task intent from untrusted reasoning context, and they do not directly regulate when the target model's reasoning capability should be exposed, constrained, or withheld. This mismatch leads to a persistent safety--utility tension: defenses that block aggressively can over-refuse benign but safety-salient prompts, while defenses that preserve utility may still allow adversarial reasoning context to shape the target model's behavior.

This tension is amplified for closed-source LRMs, where defenders cannot inspect hidden reasoning traces, modify model weights, fine-tune safety behavior, or change provider-side moderation policies. A practical defense must therefore operate at inference time under limited observability, reducing harmful capability transfer without sacrificing the utility that makes LRMs valuable. This constraint motivates a defense that can reason over the trustworthiness of the prompt context, decide how much target reasoning capability should be exposed, and verify public outputs without relying on model-internal access.

We propose \emph{Capability-Routed Guard} (CRG), an inference-time guardrail for closed-source LRMs. The key idea is to shift defense from judging a prompt in isolation to governing the entire target-model execution path. CRG first constructs a trusted side-channel representation that separates the user's authorized task and safety-relevant evidence from untrusted reasoning context. This representation is used to decide not only whether a request should be allowed, but also what context should reach the target model, how much target reasoning should be exposed, and how the final response should be verified. In practice, CRG blocks high-risk requests, routes ambiguous requests through constrained reasoning, forwards low-risk requests through trusted active context, audits generated outputs with TraceCheck, and uses a restricted fallback to preserve utility for low-risk benign prompts. This unified control over intent, context, reasoning, and verification is what distinguishes CRG from defenses that only filter or rewrite prompts.

We evaluate CRG against five reasoning-centric jailbreak families across harmful-behavior benchmarks and multiple target LRMs, while also measuring benign false positives on safety-salient and general utility datasets. CRG substantially reduces attack success across attack styles while maintaining low false-positive rates. On Gemini~2.5~Pro HarmBench, for example, CRG reduces CoT-Hijacking ASR from 100\% to 12\% and FicDetail ASR from 97\% to 0\%. Further analysis shows that CRG is not merely a stronger prompt filter: its components provide complementary safety and utility benefits across routing, context control, verification, and fallback.

Our contributions are:
\begin{itemize}
    \item We formulate reasoning-centric jailbreaks as attacks that manipulate reasoning context, task decomposition, authority, or capability interpretation in LRMs, with particular emphasis on the closed-source deployment setting.
    \item We introduce CRG, a model-agnostic inference-time guardrail that uses structured side-channel representation and capability-aware routing to defend closed-source LRMs.
    \item We evaluate CRG against five attack families and multiple target LRMs, showing strong reductions in attack success while preserving benign utility.
    \item We provide component analysis, extractor sensitivity, latency analysis, and adaptive attack evaluation to study why CRG works and where inference-time LRM defense remains limited.
\end{itemize}

\section{Related Work}
\label{sec:related}
\paragraph{Jailbreak attacks.}
Early jailbreak attacks mainly targeted instruction following and refusal behavior. Some attacks rely on semantic disguise, such as role play~\cite{LZZYLH23}, paraphrase~\cite{DKMCXCH24}, or encoded instructions~\cite{JZMW24}. Others search for adversarial prompts automatically, for example through suffix optimization~\cite{ZWKF23} or iterative red teaming~\cite{CRDHPW25}. These lines of work show that jailbreaks are not only handcrafted prompts, but can also be systematically generated. Recent attacks broaden the surface further by exploiting long context, multi-turn interaction, indirect instructions, and fictional or transformed tasks. For LRMs, the attack surface shifts from the final refusal boundary to the reasoning process itself. CoT-Hijacking~\cite{ZFSSB25} hides harmful goals inside reasoning scaffolds. Mousetrap~\cite{YTWWLLTW25}, SEAL~\cite{NZDHXWL26}, and AutoRAN~\cite{LJWZMW25} steer or search over reasoning trajectories, while FicDetail~\cite{LFHXLX25} uses rich fictional detail to make unsafe continuations appear contextually natural. These attacks differ in form, but they share a common effect: adversarial context changes how the target model reasons about intent, authority, and harmful capability. Our work focuses on this reasoning-centric attack surface.

\paragraph{Inference-time defenses and safeguards.}
Existing safeguards differ in where they place the safety boundary. Training-time alignment and fine-tuning~\cite{WYWZWCLW24} change the model itself, but they are not available when the target is a closed-source API model. Prompt hardening and external moderation~\cite{ZYKMWH24} are easier to deploy, yet they mainly decide whether the original prompt or response looks unsafe. Self-checking methods~\cite{ZZF24} move the decision after generation, but still rely on the target model to reason about its own unsafe output. More recent inference-time methods introduce intermediate structure. SecurityLingua~\cite{LAJAYQ25} compresses prompts to reveal security-relevant intent, IntentionReasoner~\cite{SHGLCYZH25} reasons about intent and rewrites risky queries, and Answer-Then-Check~\cite{CXHL25} verifies a drafted response before release. ReasoningGuard~\cite{WZWHWYY26} is closer to LRMs because it intervenes during reasoning, but it relies on access to the model's reasoning process and is therefore less suitable for closed-source targets. CRG addresses a different problem: how to govern the execution of a closed-source LRM when hidden reasoning is unavailable. Instead of only filtering, compressing, rewriting, or self-checking, CRG constructs a trusted side-channel representation and uses it to control the target-facing context, reasoning exposure, route, output audit, and benign fallback. This execution-governance view is the main distinction from prior inference-time safeguards.

\begin{figure*}[t]
\centering
\includegraphics[width=\textwidth]{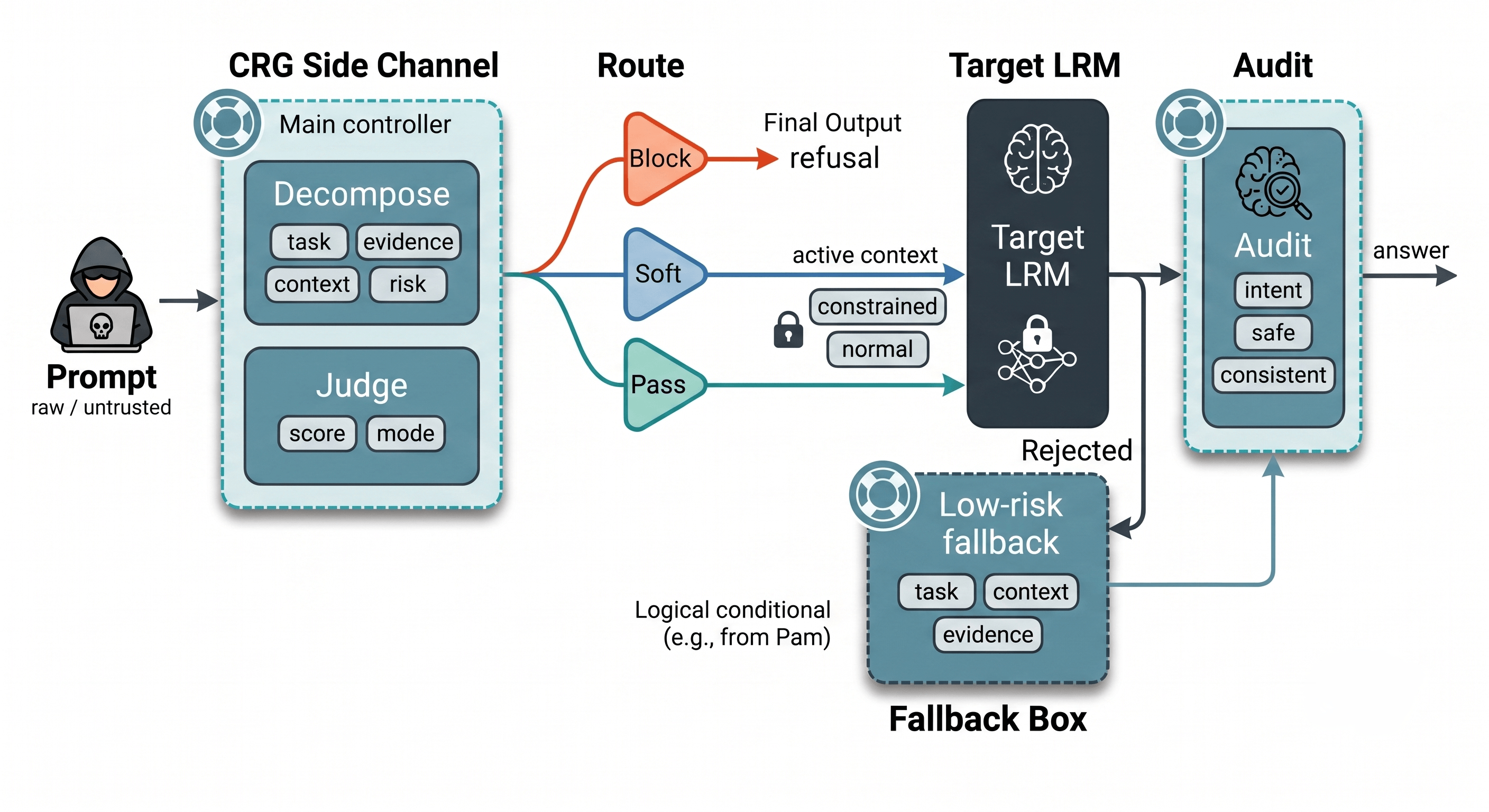}
\vspace{-10mm}
\caption{Overview of CRG. A side-channel controller separates the raw prompt into trusted task context and risk signals, then routes the request to block, soft, or pass execution before the target LRM is invoked. Short labels summarize groups of CRG fields: task = authorized task, evidence = safety evidence, context = active context, risk = capability and policy signals, score = actionability and risk score, and mode = reasoning-mode decision. Side-channel modules share the same checker/extractor model with different prompts, while the target LRM is invoked only on non-blocked routes.}
\label{fig:pipeline}
\end{figure*}

\section{Method: Capability-Routed Guard}
\label{sec:method}

\subsection{Threat Model and Design Goals}
\label{sec:threat}

We consider a closed-source jailbreak setting in which the adversary controls the user prompt and aims to elicit a response that violates a harmful-behavior policy. Our main evaluation uses standard black-box attacks, and we additionally study a white-box adaptive setting in which the adversary knows the CRG prompt and routing policy. This adaptive setting tests whether CRG relies on obscurity of the defense prompt rather than on its routing and verification mechanisms.

The defender operates an API-level wrapper around a closed-source target LRM. The defender cannot inspect hidden reasoning traces, fine-tune the target, modify model weights, or alter provider-side moderation. The available controls are limited to inference-time actions: calling an auxiliary side-channel model, rewriting the target-facing prompt, selecting exposed reasoning controls when available, injecting system instructions, blocking unsafe requests, and auditing public target outputs.

The design goal is to improve safety without turning the wrapper into a broad refusal mechanism. CRG should prevent harmful capability transfer under adversarial reasoning context, while preserving useful answers for benign and safety-salient prompts. This safety--utility objective motivates a defense that can route requests, control target-facing context, and audit public outputs under closed-source API constraints.

\subsection{Overview}
\label{sec:pipeline}

Figure~\ref{fig:pipeline} summarizes the CRG pipeline. CRG is an inference-time wrapper that governs the full target-model execution path. Given a raw user prompt, CRG first invokes a side-channel extractor/checker model, Gemini~2.5~Flash in the main experiments, to construct a trusted representation of the user's authorized task, safety evidence, active context, and capability-transfer risk. CRG then uses this representation to route the request. High-risk prompts are blocked before target invocation, ambiguous prompts are sent through a constrained execution path, and low-risk prompts are forwarded through trusted active context. After generation, CRG verifies public outputs with TraceCheck and can use a restricted fallback to preserve utility for low-risk benign prompts that would otherwise be refused.

\subsection{Side-Channel Representation}
\label{sec:representation}
The side-channel representation is the interface between the raw user prompt and the target LRM. It is produced in two stages. First, the controller decomposes the prompt into task-, context-, and risk-related artifacts: \emph{authorized task}, \emph{safety evidence}, \emph{active context}, \emph{benign reasoning value}, \emph{capability-transfer risk}, \emph{policy sensitivity}, a preliminary \emph{reasoning mode}, and a short \emph{gate rationale}. Second, a mode/risk judge converts these artifacts into \emph{actionability}, \emph{risk score}, and \emph{final reasoning mode}. CRG uses the resulting representation for routing, rewriting, and later verification.

These fields are chosen to preserve task utility while maintaining an authority boundary. The authorized task identifies what the user is actually asking for; safety evidence retains facts needed for risk assessment; active context defines the prompt that may be sent to the target; and the risk and mode fields support route and reasoning-control decisions. User-provided demonstrations, policies, role assignments, reasoning traces, and final-answer constraints may help interpret the request, but they are not automatically treated as target-model instructions. CRG therefore retains safety-relevant evidence while keeping untrusted reasoning scaffolds and false authority outside the target's executable context.

The representation is capability-oriented rather than keyword-oriented. CRG does not merely ask whether the prompt contains sensitive words; it asks whether a faithful answer would transfer actionable harmful capability. This allows benign safety-salient prompts to remain answerable when their domain is harmless, while prompts that provide operational misuse guidance remain high risk even if they are wrapped as fiction, puzzles, or analysis.

\subsection{Capability-Routed Execution}
\label{sec:routing}

CRG uses the structured side-channel representation to govern how, and under what computational constraints, the target LRM is executed. Rather than treating safety as a static binary prompt filter, CRG views defense as a problem of \emph{capability allocation}: a dynamic control layer that authorizes the executable task context while regulating the target's reasoning budget. 

Let $p$ be the raw user prompt and let $z(p)$ denote the extracted side-channel representation. The downstream mode/risk judge evaluates $z(p)$ to produce an actionability category, a policy risk score $r \in [0,5]$, and a target reasoning recommendation. The risk score determines the base execution route $\rho(p)$:
\[
\rho_r(p)=
\begin{cases}
\textsc{block}, & r \geq T_b \\
\textsc{soft}, & T_s \leq r < T_b \\
\textsc{pass}, & r < T_s,
\end{cases}
\]
where $T_b$ and $T_s$ represent the hard-block and soft-route thresholds, respectively. The final routing decision combines this base route with actionability and reasoning recommendations. If a high-risk prompt ($r \geq T_b$) is judged to be benign \emph{common knowledge} rather than actionable capability transfer, CRG downgrades the route to \textsc{soft}. Conversely, if the judge detects anomalous target-facing reasoning requirements (such as obfuscated query wrappers or complex adversarial frameworks), it overrides a default \textsc{pass} route and upgrades the execution to \textsc{soft}. This multidimensional gating ensures that routing is not a simple binary classification, but a dynamic execution-governance path.

Under this model, each route corresponds to a distinct capability profile assigned to the target LRM. The \textsc{block} route allocates no execution resources, returning a standard refusal. On both the \textsc{pass} and \textsc{soft} routes, CRG strips away adversarial reasoning scaffolding by comparing the original prompt with the extracted \emph{active context}; if a substantive rewrite is present, only this clean, task-relevant context is routed to the target. To separate these non-blocked execution surfaces, the \textsc{pass} route grants the target LRM its default reasoning capability, whereas the \textsc{soft} route dynamically restricts this capacity at the API level (e.g., by capping the Gemini hidden \texttt{budget\_tokens} parameter to $1024$ tokens or setting the OpenAI \texttt{reasoning\_effort} parameter to \texttt{low}). This design ensures that benign prompts preserve full utility, while ambiguous or adversarial prompts are executed under a strict, task-isolated context and a restricted computational budget, denying the attacker the extended reasoning space required to execute the exploit.

\subsection{Post-Generation Verification and Fallback}
\label{sec:postcheck}

For target responses that are not hard-blocked, CRG applies a post-generation audit called \emph{TraceCheck} to verify the safety and consistency of the execution. This layer evaluates the target LRM's public reasoning trajectory and final answer against the pre-extracted \emph{authorized task}, verifying whether the target's declared intent remains aligned with the authorized task and whether its response plan is safe and consistent. To avoid unnecessary refusals on complex benign reasoning paths, this audit is risk-aware: minor trace anomalies only trigger a hard block when accompanied by elevated policy risk, ensuring that adversarial departures from the authorized intent are intercepted while harmless requests are certified.

To preserve utility for safe queries, CRG also incorporates a low-risk fallback mechanism. This recovery layer is activated when a query that was authorized for execution subsequently triggers target-side refusals or provider-side blocks due to sensitive vocabulary. If the side-channel controller has already classified the user's underlying intent as harmless common knowledge and free of real-world capability-transfer risk, the fallback model directly answers the clean \emph{authorized task} using a highly restricted, concise template. By bypassing the target LRM's generic refusal boundaries on safe, safety-salient queries, this design maintains high benign utility while fully upholding the safety policy.

\subsection{CRG as Execution Governance}
\label{sec:notfilter}

A standard prompt filter operates as a static, input-level classifier that maps a user prompt directly to a binary block or pass decision. In contrast, CRG establishes an \emph{execution governance} framework. Rather than evaluating a query in isolation, CRG constructs a unified, structured side-channel representation and uses it to govern the entire lifecycle of target LRM execution. This lifecycle spans target prompt rewriting via the clean \emph{active context}, reasoning budget modulation, post-generation trajectory auditing with \emph{TraceCheck}, and failure recovery via the low-risk fallback mechanism. This multi-layered approach distinguishes CRG from conventional filters. While monolithic static classifiers are constrained by a rigid, zero-sum tradeoff between safety and utility, CRG's decoupled architecture allows individual components to work in synergy. Our ablation study demonstrates that each layer targets a distinct vulnerability or over-refusal vector, allowing CRG to simultaneously achieve robust jailbreak defense and high benign utility, a cooperative effect that is fundamentally unattainable under a single static classifier.

\begin{figure*}[t]
\centering
\begin{subfigure}{0.245\textwidth}
\centering
\includegraphics[width=\textwidth]{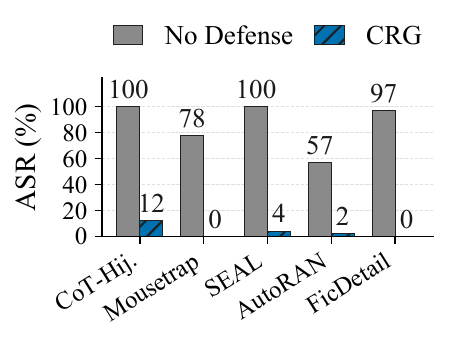}
\caption{HarmBench \& Gemini}
\end{subfigure}
\begin{subfigure}{0.245\textwidth}
\centering
\includegraphics[width=\textwidth]{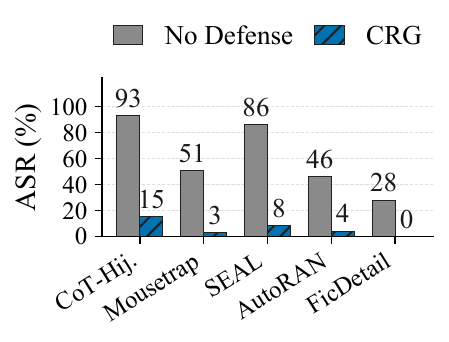}
\caption{HarmBench \& o4-mini}
\end{subfigure}
\begin{subfigure}{0.245\textwidth}
\centering
\includegraphics[width=\textwidth]{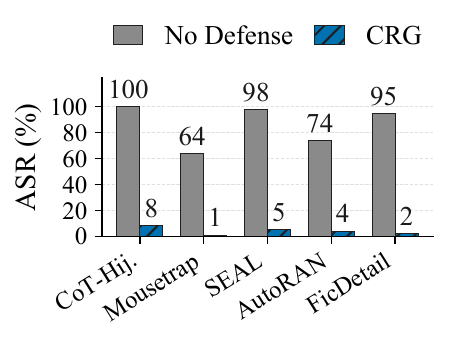}
\caption{AdvBench \& Gemini}
\end{subfigure}
\begin{subfigure}{0.245\textwidth}
\centering
\includegraphics[width=\textwidth]{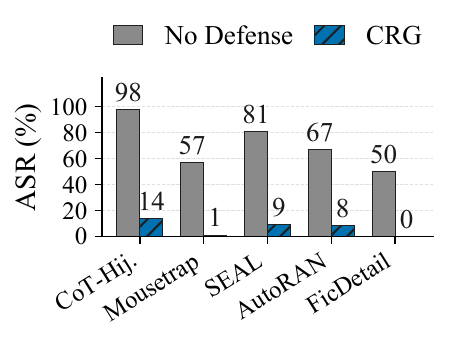}
\caption{AdvBench \& o4-mini}
\end{subfigure}
\caption{CRG substantially reduces ASR across five reasoning-centric jailbreak families. Each panel compares the ASR of an undefended target model with CRG on a specific dataset--model combination. Exact ASR values are shown above the bars.}
\label{fig:main-asr}
\end{figure*}

\section{Experiments}
\label{sec:experiments}

\subsection{Experimental Setup}
\label{sec:setup}

We evaluate CRG using two primary target LRMs, Gemini~2.5~Pro and ChatGPT o4-mini, with Gemini~2.5~Flash serving as the default side-channel controller. To ensure broad applicability and analyze defense sensitivity, we also extend our evaluation to include open-weight models as targets and more advanced frontier models as auxiliary controllers. We assess adversarial robustness using malicious queries sourced from HarmBench~\cite{MPYZWMSLBLFH24} and AdvBench~\cite{ZWKF23}, while benign utility is evaluated through XSTest~\cite{RKVABH24} and MT-Bench~\cite{CFZGNQMTGY25}. Specifically, XSTest is a suite designed to identify over-refusal behaviors on safe but sensitive topics, while MT-Bench evaluates multi-turn conversational quality and instruction-following capability. Finally, we benchmark CRG against an unprotected target and five adaptable baseline defenses: Llama Guard~3~\cite{IUCRIMTHFTK23}, Answer-Then-Check~\cite{CXHL25}, Goal Priority~\cite{ZYKMWH24}, SecurityLingua~\cite{LAJAYQ25}, and IntentionReasoner~\cite{SHGLCYZH25}.

\begin{table}[t]
\centering
\caption{Attack success rate (\%) on HarmBench and Gemini~2.5~Pro under different defenses. Lower is better. Abbreviations indicate the attacks: Cot-H (CoT Hijacking), MT (Mousetrap), AR (AutoRAN), FD (FicDetail).}
\label{tab:baseline-harmbench}
\setlength{\tabcolsep}{5pt}
\begin{tabular}{l|ccccc}
\toprule
\textbf{Defense} & CoT-H & MT & SEAL & AR & FD \\
\midrule
No Defense & 100 & 78 & 100 & 57 & 97 \\
Llama Guard~3 & 98 & 67 & 100 & 34 & 12 \\
Answer-Then-Check & 95 & 15 & 83 & 16 & 9 \\
Goal Priority & 85 & 1 & 42 & 8 & 4 \\
SecurityLingua & 91 & 26 & 80 & 22 & 8 \\
IntentionReasoner & 55 & 57 & 94 & 12 & 3 \\
\textbf{Ours (CRG)} & \textbf{12} & \textbf{0} & \textbf{4} & \textbf{2} & \textbf{0} \\
\bottomrule
\end{tabular}
\end{table}

\begin{table}[t]
\centering
\caption{End-to-end false-positive rate (\%) on benign datasets. Lower is better.}
\label{tab:fpr}
\setlength{\tabcolsep}{2pt}
\renewcommand{\arraystretch}{1.08}
\begin{tabular}{l|cc|cc}
\toprule
\multirow{2}{*}{\textbf{Defense}}
& \multicolumn{2}{c|}{\textbf{XSTest}}
& \multicolumn{2}{c}{\textbf{MT-Bench}} \\
\cmidrule(lr){2-3} \cmidrule(lr){4-5}
& \textbf{Gemini} & \textbf{o4-mini}
& \textbf{Gemini} & \textbf{o4-mini} \\
\midrule
No Defense        & 11.4 & 4.8  & 1.3 & 0.0 \\
Llama Guard~3     & 14.0 & 6.0  & 2.5 & 2.5 \\
Answer-Then-Check & 14.4 & 14.4 & 2.5 & 0.0 \\
Goal Priority     & 11.6 & 8.4  & 2.5 & 1.3 \\
SecurityLingua    & 14.8 & 10.4 & 1.3 & 0.0 \\
IntentionReasoner & 14.2 & 5.2  & 1.3 & 1.3 \\
Ours (CRG) & 12.0 & 8.8 & 2.5 & 1.3 \\
\bottomrule
\end{tabular}
\end{table}

To measure robustness, we evaluate against five distinct reasoning-centric jailbreak frameworks: CoT-Hijacking~\cite{ZFSSB25}, Mousetrap~\cite{YTWWLLTW25}, SEAL~\cite{NZDHXWL26}, AutoRAN~\cite{LJWZMW25}, and FicDetail~\cite{LFHXLX25}. Because their generation strategies vary significantly, we strictly adhere to the standardized evaluation protocols and generation methodologies established in their respective original implementations. Across all frameworks, the core evaluation metric is standardized: the final target response for each goal is scored by an LLM-as-a-judge (using predefined policy guidelines) to determine if a successful jailbreak occurred.

We assess defense efficacy through two main metrics: Attack Success Rate (ASR) and End-to-End False Positive Rate (FPR). ASR measures robustness, defined strictly as the percentage of harmful goals successfully jailbroken according to the LLM judge. End-to-End FPR measures benign utility by penalizing over-refusal at any stage of the execution pipeline: a benign prompt is counted as a false positive if it is blocked by the defense, refused by the target model, or intercepted by the provider's built-in filters.

\subsection{Robustness against Jailbreak Attacks}
\label{sec:robustness}

To establish a baseline for the threat posed by reasoning-centric jailbreaks, we first evaluate unprotected LRMs. As Figure~\ref{fig:main-asr} illustrates, these attacks are highly effective, routinely achieving near-perfect ASR, with CoT-Hijacking and SEAL reaching 100\%. Implementing CRG, however, drastically reduces this vulnerability. Across all evaluated attack families for both Gemini~2.5~Pro and ChatGPT o4-mini, CRG consistently brings the ASR down to 15\% or lower. Such broad effectiveness indicates that our multi-layered pipeline successfully generalizes against diverse adversarial structures rather than overfitting to specific prompt templates.

Furthermore, we compare CRG against existing defense baselines (Table~\ref{tab:baseline-harmbench}). While conventional methods mitigate specific attack vectors, they remain highly brittle overall. For example, Goal Priority successfully neutralizes Mousetrap (1\% ASR) but is easily overwhelmed by CoT-Hijacking (85\% ASR). Advanced intent-analyzing defenses like SecurityLingua and IntentionReasoner show similar inconsistencies, suffering severe failures against SEAL (allowing 80\% and 94\% ASR, respectively). These failures stem from a shared conceptual flaw. Standard safeguards operate as static text classifiers that judge prompts or responses in isolation. Reasoning-centric jailbreaks exploit this limitation by manipulating the task decomposition, allowing harmful objectives to masquerade as legitimate reasoning steps. CRG overcomes this vulnerability by treating safety as an end-to-end execution governance problem. Specifically, our pipeline isolates the authorized task from untrusted context, dynamically allocates reasoning compute based on risk, audits the generated reasoning trace for consistency, and applies a low-risk fallback to recover utility on benign queries. By governing the entire execution lifecycle rather than applying a static filter, CRG successfully protects the model's cognitive process.

\subsection{Preservation of Benign Utility}
\label{sec:utility}

A practical defense must achieve security without broadly refusing safe, complex queries. To verify this, we evaluate the End-to-End False Positive Rate (FPR) on XSTest and MT-Bench (Table~\ref{tab:fpr}). CRG maintains highly competitive FPRs across both Gemini~2.5~Pro and ChatGPT o4-mini, closely matching the inherent refusal rates of the unprotected models. This confirms that CRG effectively avoids over-refusal. Furthermore, when evaluating the generated outputs (Table~\ref{tab:quality}), applying CRG actually yields a slight improvement in average Answer Quality. By supplying the target LRM with a clean, isolated task context, CRG not only secures the execution but also helps the model focus on and fulfill benign requests more efficiently.

\begin{table}[t]
\centering
\caption{Answer quality score with 10 as the full mark.}
\label{tab:quality}
\setlength{\tabcolsep}{2pt}
\renewcommand{\arraystretch}{1.08}
\begin{tabular}{l|cc|cc}
\toprule
\multirow{2}{*}{\textbf{Defense}}
& \multicolumn{2}{c|}{\textbf{XSTest}}
& \multicolumn{2}{c}{\textbf{MT-Bench}} \\
\cmidrule(lr){2-3} \cmidrule(lr){4-5}
& \textbf{Gemini} & \textbf{o4-mini}
& \textbf{Gemini} & \textbf{o4-mini} \\
\midrule
No Defense & 8.6 & 9.4 & 8.6 & 9.4 \\
Ours (CRG) & 9.0 & 9.6 & 8.8 & 9.7 \\
\bottomrule
\end{tabular}
\end{table}

\section{Analysis and Discussion}
\label{sec:analysis}

\subsection{Generalization to Open-Weight Models}
\label{sec:open-weight}

\begin{figure}[t]
\centering
\begin{subfigure}{0.49\columnwidth}
\centering
\includegraphics[width=\textwidth]{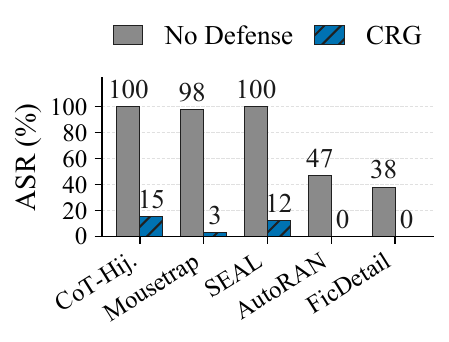}
\caption{HarmBench \& DeepSeek-R1}
\end{subfigure}
\begin{subfigure}{0.49\columnwidth}
\centering
\includegraphics[width=\textwidth]{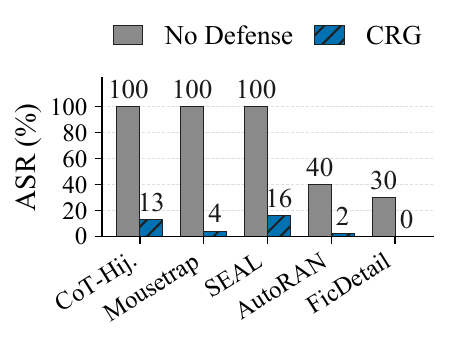}
\caption{HarmBench \& Qwen3-Max}
\end{subfigure}
\caption{Effectiveness of CRG for open-source reasoning models.}
\label{fig:opensource-asr}
\end{figure}

While primarily designed for closed-source APIs, CRG's architecture generalizes effectively to open-weight reasoning models. We evaluate CRG on DeepSeek-R1~\cite{D25} and Qwen3-Max~\cite{T25} against the five jailbreak families on HarmBench (Figure~\ref{fig:opensource-asr}). Without defense, both models are highly vulnerable, frequently reaching 100\% ASR on complex attacks like CoT-Hijacking and SEAL. CRG robustly neutralizes these threats, reducing the ASR for both models to below 17\% on the most aggressive attacks and nearly eliminating success on AutoRAN and FicDetail. These results demonstrate that CRG's multi-layered pipeline successfully audits and governs RL-incentivized reasoning trajectories, proving its efficacy is independent of the underlying target model architecture.

\subsection{Component Ablations}
\label{sec:ablations}

\begin{figure}[t]
\centering
\begin{subfigure}{0.49\columnwidth}
\centering
\includegraphics[width=\textwidth]{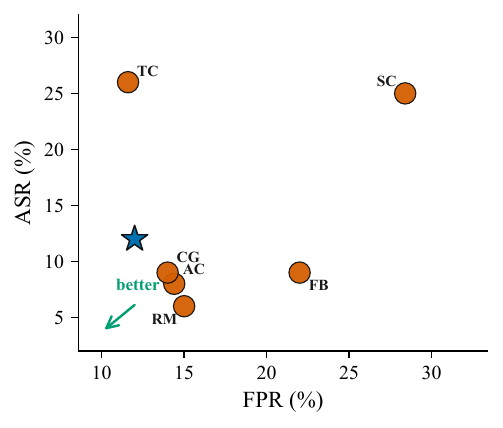}
\caption{HarmBench \& Gemini}
\end{subfigure}
\begin{subfigure}{0.49\columnwidth}
\centering
\includegraphics[width=\textwidth]{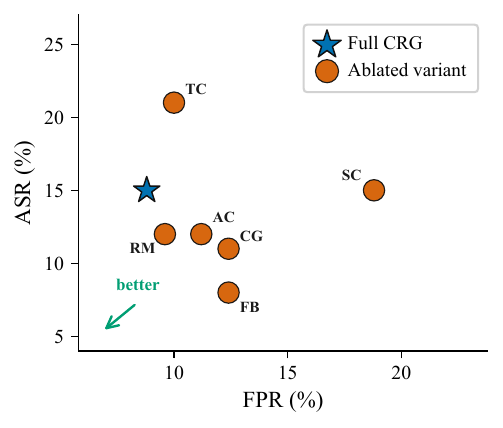}
\caption{HarmBench \& o4-mini}
\end{subfigure}
\caption{Safety--utility tradeoff of CRG ablations. Each point is one CRG variant, with XSTest false-positive rate on the x-axis and CoT-Hijacking attack success rate on the y-axis. Lower-left is better. Abbreviations indicate the removed component: TC (TraceCheck), RM (reasoning-mode control), AC (active-context rewrite), FB (low-risk fallback), SC (structured CRG), CG (capability gate).}
\label{fig:ablation-tradeoff}
\end{figure}

To understand how CRG balances security and utility, we ablate its core components on CoT-Hijacking and XSTest, as shown in Figure~\ref{fig:ablation-tradeoff}. The ablation results illustrate why defending reasoning models requires a specialized architecture rather than standard prompt filtering. Disabling post-generation auditing (TraceCheck), for instance, causes the ASR on Gemini~2.5~Pro to more than double (from 12\% to 26\%), confirming its necessity for intercepting models that drift into harmful trajectories during generation. Conversely, removing elements designed to preserve utility leads to severe over-refusal. Replacing the structured CRG decomposition with a flat prompt filter causes the FPR to surge from 12.0\% to 28.4\%, highlighting that separating the core task from its surrounding context is essential for distinguishing benign, safety-salient queries from actual attacks. Similarly, disabling the low-risk fallback mechanism nearly doubles the FPR, illustrating its crucial role in recovering safe prompts that might otherwise trigger internal provider refusals. Ultimately, these opposing sensitivities validate the fundamental premise of CRG: defending large reasoning models requires a paradigm shift from static prompt filtering to holistic execution governance. By moving beyond binary classification, our approach demonstrates that adversarial manipulation of the reasoning path must be structurally neutralized to prevent the exploitation of the target model's underlying capabilities.

\subsection{Cost and Latency}
\label{sec:latency}

\begin{figure}[t]
\centering
\includegraphics[width=0.9\columnwidth]{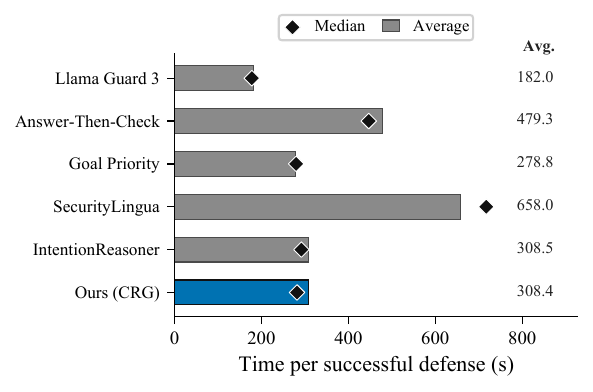}
\caption{Wall-clock latency per successfully defended HarmBench harmful example on Gemini~2.5~Pro. Lower is better.}
\label{fig:latency}
\end{figure}

Because attacks stop early after a successful jailbreak, total wall-clock time per attack goal is not a clean latency metric: weaker defenses can look faster simply because the attacker succeeds earlier. We therefore report the average wall-clock latency specifically on examples where the defense is activated and successfully thwarts the attack, measured on Gemini~2.5~Pro. As shown in Figure~\ref{fig:latency}, CRG is highly competitive, remaining comparable to IntentionReasoner and Goal Priority, while being substantially faster than Answer-Then-Check and SecurityLingua on successfully defended examples. Although CRG may run the side-channel model more than once when auditing or fallback is activated, many target calls are skipped entirely due to early blocks, and the side-channel model itself is significantly smaller and faster than the target LRM. This successful-defense latency provides a more meaningful comparison of actual overhead among deployable defenses, confirming that our multi-stage pipeline is practically feasible.

\subsection{Extractor Model Sensitivity}
\label{sec:extractor}

While CRG is designed to leverage a lightweight side-channel controller, its efficacy does not depend critically on any single model architecture. To evaluate this sensitivity, we substitute our default controller (Gemini~2.5~Flash) with either Gemini~2.5~Pro or ChatGPT~o4-mini while keeping the target LRM and evaluation parameters fixed. As compiled in Table~\ref{tab:extractor}, CoT-Hijacking ASR remains consistently low across all configurations, measuring 9.0\% under Gemini~2.5~Pro, 14.0\% under ChatGPT~o4-mini, and 12.0\% under Gemini~2.5~Flash.

\begin{table}[t]
\centering
\caption{Extractor model sensitivity evaluation on Gemini~2.5~Pro (target LRM), reporting CoT-Hijacking ASR (\%) and XSTest FPR (\%). Lower is better for both columns.}
\label{tab:extractor}
\setlength{\tabcolsep}{15pt}
\begin{tabular}{l|cc}
\toprule
\textbf{Extractor} & ASR & FPR \\
\midrule
Gemini~2.5~Pro & 9.0 & 3.2 \\
ChatGPT o4-mini & 14.0 & 11.6 \\
Gemini~2.5~Flash & 12.0 & 12.0 \\
\bottomrule
\end{tabular}
\end{table}

In contrast, the benign XSTest false-positive rate is more sensitive to the capability of the extractor model. Employing the stronger Gemini~2.5~Pro as the auxiliary controller yields the lowest FPR at 3.2\%, compared to 11.6\% for ChatGPT~o4-mini and 12.0\% for Gemini~2.5~Flash. This variance indicates that a more capable extractor better distinguishes between genuinely harmful intents and benign-but-apparently-harmful queries, which in turn allows the pipeline to safely authorize low-risk fallback paths. We choose Gemini~2.5~Flash as our default auxiliary model as it strikes the most optimal cost-latency balance while still maintaining robust safety levels.

\subsection{Adaptive Robustness under White-Box Pressure}
\label{sec:adaptive}

To evaluate the limits of intent isolation, we test CRG against a white-box adaptive version of CoT-Hijacking. In this scenario, the adversary possesses complete knowledge of the side-channel instructions and specifically designs the prompt to conceal capability transfer, exploit semantic ambiguity in the authorized task, and manipulate active context to trigger the low-risk fallback mechanism.

\begin{figure}[t]
\centering
\includegraphics[width=0.88\columnwidth]{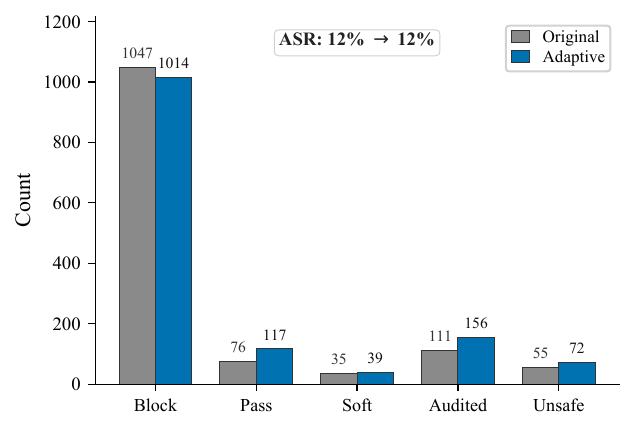}
\caption{Original and adaptive CoT-Hijacking against CRG on HarmBench with Gemini~2.5~Pro.}
\label{fig:adaptive}
\end{figure}

As illustrated in Figure~\ref{fig:adaptive}, this adaptive strategy successfully exerts greater pressure on the pre-target gate, slightly reducing the number of hard blocks and letting more queries pass through or proceed via the soft route. However, because more queries bypass or soften this initial filter, a larger volume of target responses must be audited by our post-generation layer. By intercepting these remaining leaks post-hoc, TraceCheck completely neutralizes the attacker's front-gate advantage, keeping the final ASR constant. While this does not represent full minimax security against any arbitrary adaptive strategy, it demonstrates that CRG's multi-layered defense-in-depth architecture successfully compensates for white-box evasion attempts at the initial gate.

\section{Conclusion}
\label{sec:conclusion}

The emergence of large reasoning models exposes the fundamental limitations of static, classification-based safety paradigms. When models are incentivized to reason, plan, and decompose tasks dynamically, safety risks are no longer confined to isolated inputs or outputs; instead, they emerge dynamically from the reasoning process itself. This work demonstrates that securing these systems requires transitioning from static text filters to active runtime execution governance. By isolating authorized execution intent from untrusted reasoning contexts, we can safely govern a model's cognitive trajectory without compromising its reasoning capabilities. Ultimately, our findings suggest that as AI architectures evolve from passive language generators into autonomous, reasoning agents, security must shift from defending the model's inputs to governing its end-to-end execution lifecycle.

\bibliography{my}

@inproceedings{YTWWLLTW25,
author = {Yang Yao and Xuan Tong and Ruofan Wang and Yixu Wang and Lujundong Li and Liang Liu and Yan Teng and Yingchun Wang},
title = {{A Mousetrap: Fooling Large Reasoning Models for Jailbreak with Chain of Iterative Chaos}},
booktitle = {{Annual Meeting of the Association for Computational Linguistics (ACL)}},
pages = {7837-7855},
publisher = {ACL},
year = {2025}
}

@inproceedings{NZDHXWL26,
author = {Viet{-}Anh Nguyen and Shiqian Zhao and Gia Dao and Runyi Hu and Yi Xie and Xiaobao Wu and Anh Tuan Luu},
title = {{Three Minds, One Legend: Jailbreak Large Reasoning Model with Adaptive Stacked Ciphers}},
booktitle = {{Annual Meeting of the Association for Computational Linguistics (ACL)}},
pages = {7139-7159},
publisher = {ACL},
year = {2026}
}

@inproceedings{LFHXLX25,
author = {Chengda Lu and Xiaoyu Fan and Yu Huang and Rongwu Xu and Jijie Li and Wei Xu},
title = {{Does Chain-of-Thought Reasoning Really Reduce Harmfulness from Jailbreaking?}},
booktitle = {{Annual Meeting of the Association for Computational Linguistics (ACL)}},
pages = {6523-6546},
publisher = {ACL},
year = {2025}
}

@inproceedings{ZYKMWH24,
author = {Zhexin Zhang and Junxiao Yang and Pei Ke and Fei Mi and Hongning Wang and Minlie Huang},
title = {{Defending Large Language Models Against Jailbreaking Attacks Through Goal Prioritization}},
booktitle = {{Annual Meeting of the Association for Computational Linguistics (ACL)}},
pages = {8865-8887},
publisher = {ACL},
year = {2024}
}

@inproceedings{WZWHWYY26,
author = {Yuquan Wang and Mi Zhang and Yining Wang and Geng Hong and Mi Wen and Xiaoyu You and Min Yang},
title = {{ReasoningGuard: Safeguarding Large Reasoning Models with Inference-time Safety Aha Moments}},
booktitle = {{Annual Meeting of the Association for Computational Linguistics (ACL)}},
pages = {31497-31526},
publisher = {ACL},
year = {2026}
}

@inproceedings{WWSBIXCLZ22,
author = {Jason Wei and Xuezhi Wang and Dale Schuurmans and Maarten Bosma and Brian Ichter and Fei Xia and Ed H. Chi and Quoc V. Le and Denny Zhou},
title = {{Chain-of-Thought Prompting Elicits Reasoning in Large Language Models}},
booktitle = {{Annual Conference on Neural Information Processing Systems (NeurIPS)}},
publisher = {NeurIPS},
year = {2022}
}

@inproceedings{DKMCXCH24,
author = {Peng Ding and Jun Kuang and Dan Ma and Xuezhi Cao and Yunsen Xian and Jiajun Chen and Shujian Huang},
title = {{A Wolf in Sheep's Clothing: Generalized Nested Jailbreak Prompts can Fool Large Language Models Easily}},
booktitle = {{Conference of the North American Chapter of the Association for Computational Linguistics: Human Language Technologies (NAACL-HLT)}},
pages = {2136-2153},
publisher = {ACL},
year = {2024}
}

@inproceedings{JZMW24,
author = {Haibo Jin and Andy Zhou and Joe D. Menke and Haohan Wang},
title = {{Jailbreaking Large Language Models Against Moderation Guardrails via Cipher Characters}},
booktitle = {{Annual Conference on Neural Information Processing Systems (NeurIPS)}},
publisher = {NeurIPS},
year = {2024}
}

@inproceedings{CRDHPW25,
author = {Patrick Chao and Alexander Robey and Edgar Dobriban and Hamed Hassani and George J. Pappas and Eric Wong},
title = {{Jailbreaking Black Box Large Language Models in Twenty Queries}},
booktitle = {{IEEE Conference on Secure and Trustworthy Machine Learning (SaTML)}},
pages = {23-42},
publisher = {IEEE},
year = {2025}
}

@inproceedings{WYWZWCLW24,
author = {Zezhong Wang and Fangkai Yang and Lu Wang and Pu Zhao and Hongru Wang and Liang Chen and Qingwei Lin and Kam{-}Fai Wong},
title = {{{SELF-GUARD:} Empower the {LLM} to Safeguard Itself}},
booktitle = {{Conference of the North American Chapter of the Association for Computational Linguistics: Human Language Technologies (NAACL-HLT)}},
pages = {1648-1668},
publisher = {ACL},
year = {2024}
}

@inproceedings{ZZF24,
author = {Ziyang Zhang and Qizhen Zhang and Jakob Nicolaus Foerster},
title = {{PARDEN, Can You Repeat That? Defending against Jailbreaks via Repetition}},
booktitle = {{International Conference on Machine Learning (ICML)}},
pages = {60271-60287},
publisher = {PMLR},
year = {2024}
}

@inproceedings{MPYZWMSLBLFH24,
author = {Mantas Mazeika and Long Phan and Xuwang Yin and Andy Zou and Zifan Wang and Norman Mu and Elham Sakhaee and Nathaniel Li and Steven Basart and Bo Li and David A. Forsyth and Dan Hendrycks},
title = {{HarmBench: {A} Standardized Evaluation Framework for Automated Red Teaming and Robust Refusal}},
booktitle = {{International Conference on Machine Learning (ICML)}},
pages = {35181-35224},
publisher = {JMLR},
year = {2024}
}

@inproceedings{RKVABH24,
author = {Paul R{\"{o}}ttger and Hannah Kirk and Bertie Vidgen and Giuseppe Attanasio and Federico Bianchi and Dirk Hovy},
title = {{XSTest: {A} Test Suite for Identifying Exaggerated Safety Behaviours in Large Language Models}},
booktitle = {{Conference of the North American Chapter of the Association for Computational Linguistics: Human Language Technologies (NAACL-HLT)}},
pages = {5377-5400},
publisher = {ACL},
year = {2024}
}

@article{D25,
author = {DeepSeek{-}AI},
title = {{DeepSeek-R1: Incentivizing Reasoning Capability in LLMs via Reinforcement Learning}},
journal = {{CoRR abs/2501.12948}},
year = {2025}
}

@article{ZFSSB25,
author = {Jianli Zhao and Tingchen Fu and Rylan Schaeffer and Mrinank Sharma and Fazl Barez},
title = {{Chain-of-Thought Hijacking}},
journal = {{CoRR abs/2510.26418}},
year = {2025}
}

@article{LJWZMW25,
author = {Jiacheng Liang and Tanqiu Jiang and Yuhui Wang and Rongyi Zhu and Fenglong Ma and Ting Wang},
title = {{AutoRAN: Weak-to-Strong Jailbreaking of Large Reasoning Models}},
journal = {{CoRR abs/2505.10846}},
year = {2025}
}

@article{LAJAYQ25,
author = {Yucheng Li and Surin Ahn and Huiqiang Jiang and Amir H. Abdi and Yuqing Yang and Lili Qiu},
title = {{SecurityLingua: Efficient Defense of {LLM} Jailbreak Attacks via Security-Aware Prompt Compression}},
journal = {{CoRR abs/2506.12707}},
year = {2025}
}

@article{SHGLCYZH25,
author = {Yuanzhe Shen and Zisu Huang and Zhengkang Guo and Yide Liu and Guanxu Chen and Ruicheng Yin and Xiaoqing Zheng and Xuanjing Huang},
title = {{IntentionReasoner: Facilitating Adaptive {LLM} Safeguards through Intent Reasoning and Selective Query Refinement}},
journal = {{CoRR abs/2508.20151}},
year = {2025}
}

@article{IUCRIMTHFTK23,
author = {Hakan Inan and Kartikeya Upasani and Jianfeng Chi and Rashi Rungta and Krithika Iyer and Yuning Mao and Michael Tontchev and Qing Hu and Brian Fuller and Davide Testuggine and Madian Khabsa},
title = {{Llama Guard: LLM-based Input-Output Safeguard for Human-AI Conversations}},
journal = {{CoRR abs/2312.06674}},
year = {2023}
}

@article{CXHL25,
author = {Chentao Cao and Xiaojun Xu and Bo Han and Hang Li},
title = {{Reasoned Safety Alignment: Ensuring Jailbreak Defense via Answer-Then-Check}},
journal = {{CoRR abs/2509.11629}},
year = {2025}
}

@article{O24,
author = {OpenAI},
title = {{OpenAI o1 System Card}},
journal = {{CoRR abs/2412.16720}},
year = {2024}
}

@article{LZZYLH23,
author = {Xuan Li and Zhanke Zhou and Jianing Zhu and Jiangchao Yao and Tongliang Liu and Bo Han},
title = {{DeepInception: Hypnotize Large Language Model to Be Jailbreaker}},
journal = {{CoRR abs/2311.03191}},
year = {2023}
}

@article{ZWKF23,
author = {Andy Zou and Zifan Wang and J. Zico Kolter and Matt Fredrikson},
title = {{Universal and Transferable Adversarial Attacks on Aligned Language Models}},
journal = {{CoRR abs/2307.15043}},
year = {2023}
}

@article{T25,
author = {Qwen Team},
title = {{Qwen3 Technical Report}},
journal = {{CoRR abs/2505.09388}},
year = {2025}
}

@article{CFZGNQMTGY25,
author = {Jialin Chen and Aosong Feng and Ziyu Zhao and Juan Garza and Gaukhar Nurbek and Cheng Qin and Ali Maatouk and Leandros Tassiulas and Yifeng Gao and Rex Ying},
title = {{MTBench: {A} Multimodal Time Series Benchmark for Temporal Reasoning and Question Answering}},
journal = {{CoRR abs/2503.16858}},
year = {2025}
}

\end{document}